\documentclass[a4paper,11pt]{article}
\pdfoutput=1 

\usepackage{jheppub} 

\title{\boldmath Search for $CP$ Violation in the Decay $D^+\rightarrow K^0_S K^+$}
\collaboration{Belle Collaboration}
  \author[21]{B.~R.~Ko} 
  \author[21]{E.~Won} 
  \author[9]{I.~Adachi} 
  \author[51]{H.~Aihara} 
  \author[3]{K.~Arinstein} 
  \author[41]{D.~M.~Asner} 
  \author[16]{T.~Aushev} 
  \author[45]{A.~M.~Bakich} 
  \author[14]{K.~Belous} 
  \author[33]{V.~Bhardwaj} 
% ...

% ...
\affiliation[3]{Budker Institute of Nuclear Physics SB RAS and Novosibirsk State University, Novosibirsk 630090, Russian Federation}
% ...
\affiliation[9]{High Energy Accelerator Research Organization (KEK), Tsukuba 305-0801, Japan}
% ...
\affiliation[14]{Institute for High Energy Physics, Protvino 142281, Russian Federation}
% ...
\affiliation[16]{Institute for Theoretical and Experimental Physics, Moscow 117218, Russian Federation}
% ...
\affiliation[21]{Korea University, Seoul 136-713, South Korea}
% ...
\affiliation[33]{Nara Women's University, Nara 630-8506, Japan}
% ...
\affiliation[41]{Pacific Northwest National Laboratory, Richland, WA 99352, USA}
% ...
\affiliation[45]{School of Physics, University of Sydney, NSW 2006, Australia}
% ...
\affiliation[51]{Department of Physics, University of Tokyo, Tokyo 113-0033, Japan}
% ...

\abstract{

We search for $CP$ violation in the decay $D^+\rightarrow K^0_S K^+$
using a data sample with an integrated luminosity of 977 fb$^{-1}$
collected with the Belle detector at the KEKB $e^+e^-$
asymmetric-energy collider. No $CP$ violation has been observed and
the $CP$ asymmetry in $D^+\rightarrow K^0_S K^+$ decay is measured to
be $(-0.25\pm0.28\pm0.14)\%$, which is the most sensitive measurement
to date. After subtracting $CP$ violation due to $K^0-\bar{K}^0$
mixing, the $CP$ asymmetry in $D^+\rightarrow\bar{K}^0 K^+$ decay is
found to be $(+0.08\pm0.28\pm0.14)\%$.
}

\begin{document} 
\maketitle
\flushbottom

\section{Introduction}\label{SEC:INTRODUCTION}
Studies of $CP$ violation in charmed meson decays provide a promising
opportunity to search for new physics beyond the standard model
(SM)~\cite{CICERONE} in the absence of disagreement between experimental
measurements and the SM interpretation of $CP$ violation in $K$ and
$B$ meson decays~\cite{PDG2012,HFAG,CKM2010}.
Recently, the LHCb collaboration has reported $\Delta
A_{CP}=(-0.82\pm0.21\pm0.11)\%$~\cite{DELTACPV_LHCB} where $\Delta
A_{CP}$ is the $CP$ asymmetry difference between $D^0\rightarrow
K^+K^-$~\footnote{Throughout this paper, the charge-conjugate decay
  modes are implied unless stated otherwise.} and
$D^0\rightarrow\pi^+\pi^-$ decays. Thereafter, the CDF collaboration
has also announced $\Delta
A_{CP}=(-0.62\pm0.21\pm0.10)\%$~\cite{DELTACPV_CDF}, which strongly
supports the non-zero $\Delta A_{CP}$ measured from the LHCb
collaboration. Together with results from the BaBar and Belle
collaborations, the value of $\Delta A_{CP}$ is significantly
different from zero~\cite{HFAG_PART}. Taking into account that the
indirect $CP$ asymmetries in the two decays are approximately
equal~\cite{YAY}, $\Delta A_{CP}$ can be expressed as
\begin{equation}
  \Delta A_{CP}=\Delta a^{\rm dir}_{CP} + a^{\rm ind}_{CP}\Delta\langle t \rangle/\tau,
  \label{EQ:DACP}
\end{equation}
where $a^{\rm dir}_{CP}$ and $a^{\rm ind}_{CP}$ denote direct and
indirect $CP$ violation, respectively, and $\langle t\rangle/\tau$ is
the mean proper decay time of the selected signal sample in units of
the $D^0$ lifetime~\cite{HFAG_DACP}. The factor $\Delta\langle
t\rangle/\tau$ in eq.~(\ref{EQ:DACP}) depends on the experimental
conditions and the largest value reported to date is $0.26\pm0.01$
from the CDF measurement~\cite{DELTACPV_CDF}. Therefore, $\Delta A_{CP}$
reveals a significant direct $CP$ violation difference between the two
decays. Within the SM, direct $CP$ violation in the charm sector is
expected to be present only in singly Cabibbo-suppressed (SCS) decays,
and even there is expected to be small,
$\mathcal{O}(0.1\%)$~\cite{SMCP}. Hence, the current $\Delta A_{CP}$
measurements engender questions of whether the origin of the asymmetry
lies within~\cite{WSM1,WSM2,WSM3,WSM4} or
beyond~\cite{BSM1,BSM2,BSM3,BSM4} the SM. The origin of $\Delta
A_{CP}$ calls for the precise measurements of $A_{CP}$ in
$D^0\rightarrow K^+K^-$ and $D^0\rightarrow\pi^+\pi^-$. A
complementary test is a precise measurement of $A_{CP}$ in another SCS
charmed hadron decay, $D^+\rightarrow\bar{K}^0 K^+$, as suggested in
ref.~\cite{WSM3}.
\begin{figure}[htbp]
\centering
\mbox{
  \includegraphics[height=0.21\textwidth,width=0.47\textwidth]{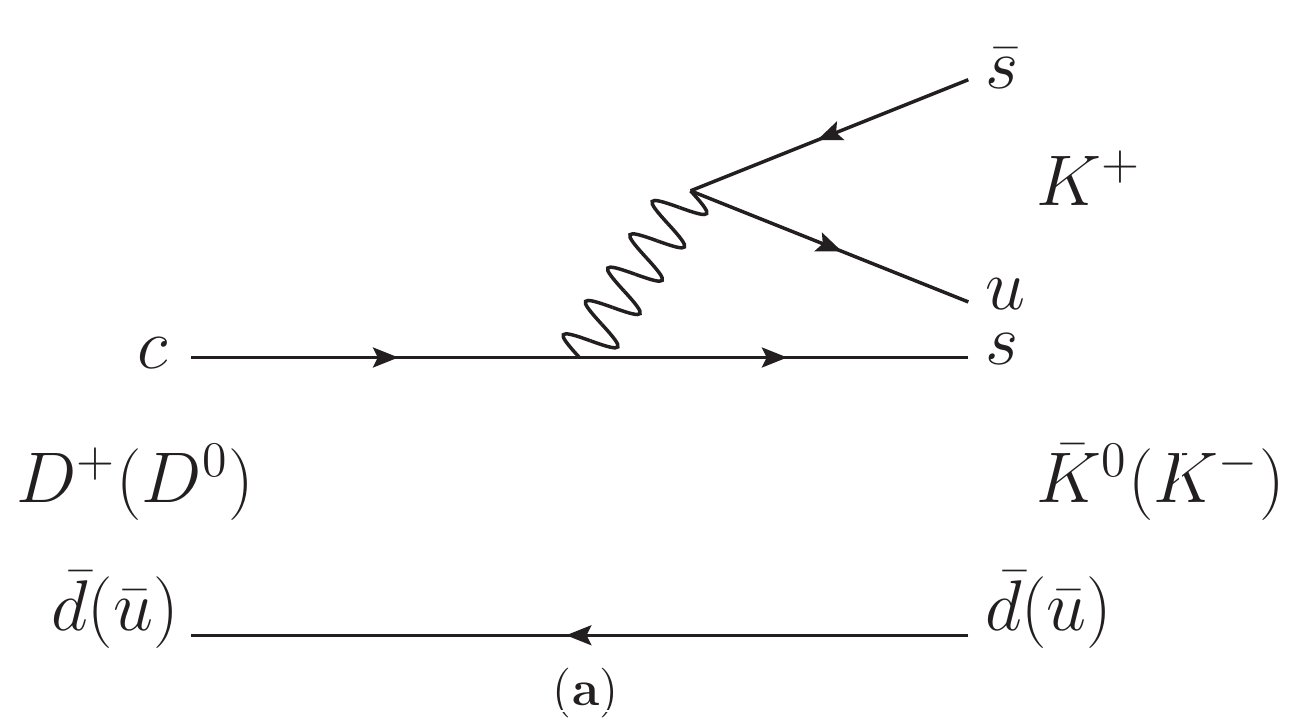}
}
\mbox{
  \includegraphics[height=0.21\textwidth,width=0.47\textwidth]{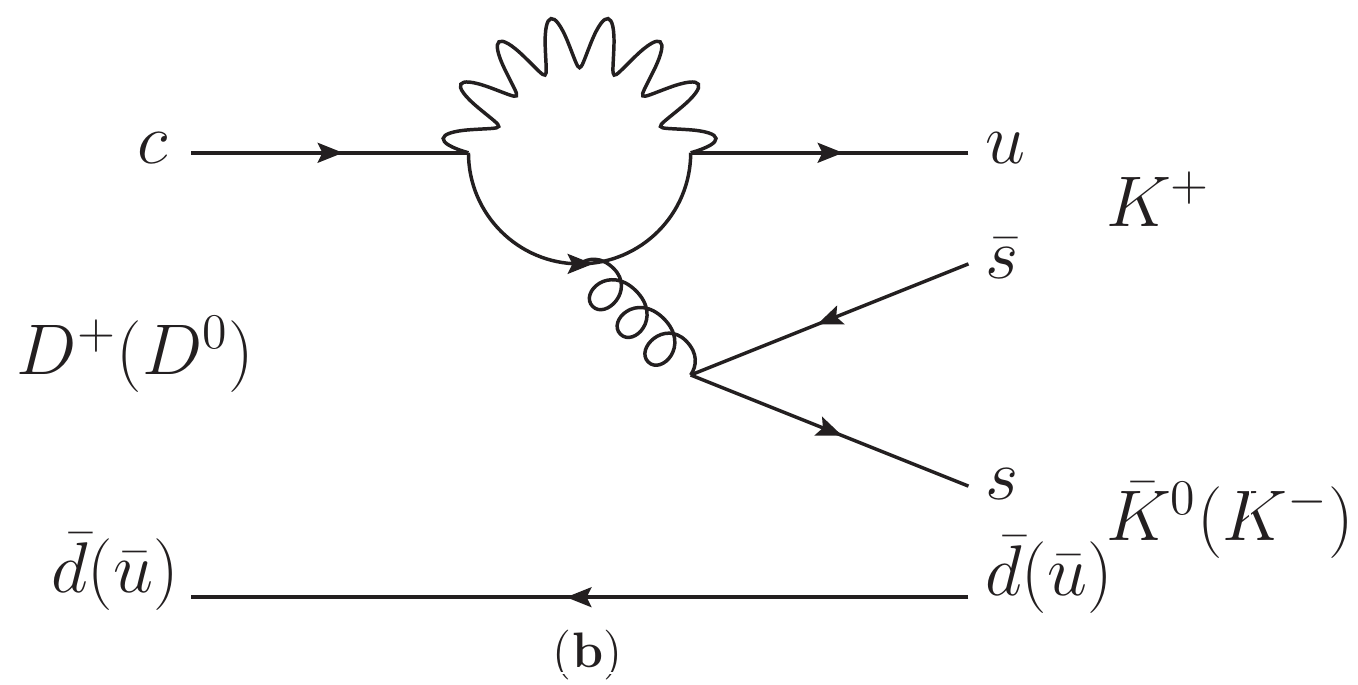}
}
\mbox{
  \includegraphics[height=0.21\textwidth,width=0.42\textwidth]{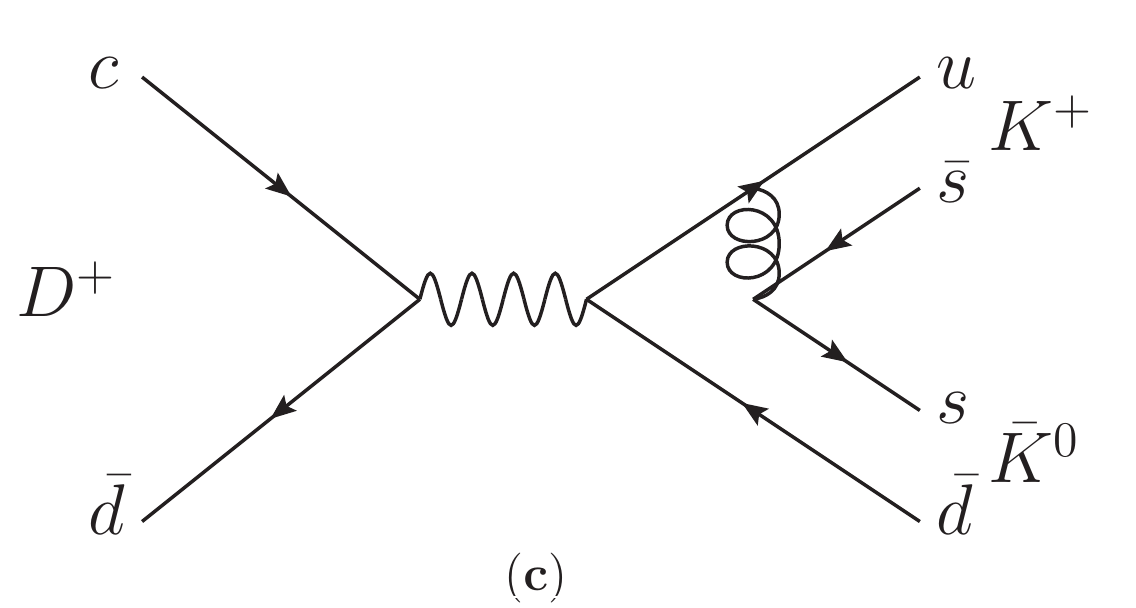}
}
\mbox{
  \includegraphics[height=0.21\textwidth,width=0.42\textwidth]{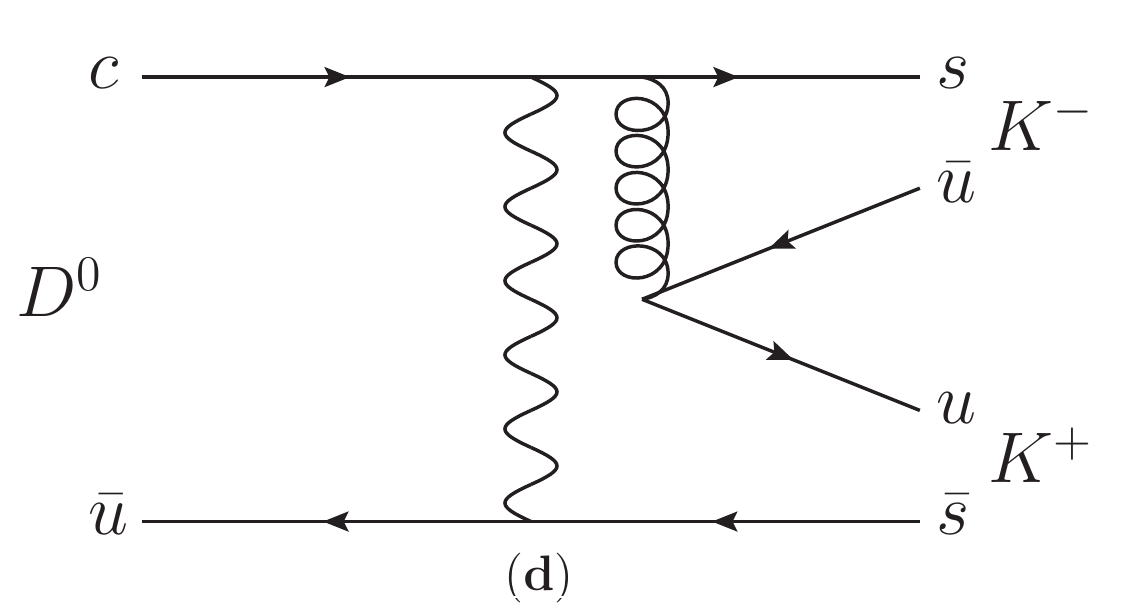}
}
\caption{Feynman diagrams of $D^+\rightarrow\bar{K}^0 K^+$ and
  $D^0\rightarrow K^+K^-$ decays.}
\label{FIG:DIAGRAMS}
\end{figure}
As shown in figures~\ref{FIG:DIAGRAMS}(a) and~\ref{FIG:DIAGRAMS}(b),
the decay $D^+\rightarrow\bar{K}^0 K^+$ shares the same decay diagrams
with $D^0\rightarrow K^+K^-$ by exchanging the spectator quarks,
$d\leftrightarrow u$. Although there are additional contributions to
the two decays as shown in figures~\ref{FIG:DIAGRAMS}(c)
and~\ref{FIG:DIAGRAMS}(d), these are expected to be small due to
helicity- and color-suppression considerations~\footnote{In helicity
  suppression, a spinless meson decaying to a back-to-back
  quark-antiquark pair is suppressed by the conservation of angular
  momentum. In color suppression, the final state quarks are required
  to carry the correct color charge in order for the final state to be
  colorless.}. Therefore, neglecting the latter contributions in
$D^+\rightarrow\bar{K}^0 K^+$ and $D^0\rightarrow K^+K^-$ decays, the
direct $CP$ asymmetries in the two decays are expected to be the same.

In this paper, we report results from a search for $CP$ violation in
the decay $D^+\rightarrow K^0_S K^+$ that originates from
$D^+\rightarrow\bar{K}^0 K^+$ decay, where $K^0_S$ decays to
$\pi^+\pi^-$. The $CP$ asymmetry in the decay, $A_{CP}$, is then
defined as
\begin{eqnarray}
  \nonumber
  A^{D^+\rightarrow K^0_S K^+}_{CP}
  &\equiv&\frac
  {\Gamma(D^+\rightarrow\bar{K}^0 K^+)\Gamma(\bar{K}^0\rightarrow\pi^+\pi^-)-\Gamma(D^-\rightarrow K^0 K^-)\Gamma(K^0\rightarrow\pi^+\pi^-)}
  {\Gamma(D^+\rightarrow\bar{K}^0 K^+)\Gamma(\bar{K}^0\rightarrow\pi^+\pi^-)+\Gamma(D^-\rightarrow K^0 K^-)\Gamma(K^0\rightarrow\pi^+\pi^-)}\\
  &=&\frac{A^{D^+\rightarrow\bar{K}^0 K^+}_{CP} + A^{\bar{K}^0}_{CP}}{1+A^{D^+\rightarrow\bar{K}^0 K^+}_{CP}A^{\bar{K}^0}_{CP}}
  \simeq A^{D^+\rightarrow\bar{K}^0 K^+}_{CP} + A^{\bar{K}^0}_{CP},
  \label{EQ:ACP}
\end{eqnarray}
where $\Gamma$ is the partial decay width. In eq.~(\ref{EQ:ACP}),
$A^{D^+\rightarrow\bar{K}^0K^+}_{CP}$ is the $CP$ asymmetry in the
decay $D^+\rightarrow\bar{K}^0K^+$ and $A^{\bar{K}^0}_{CP}$ is that in
$\bar{K}^0\rightarrow\pi^+\pi^-$ decay induced by $K^0-\bar{K}^0$
mixing in the SM~\cite{K0CP1,K0CP2,K0CP3} in which the decay
$\bar{K}^0\rightarrow\pi^+\pi^-$ arises from
$K^0_S\rightarrow\pi^+\pi^-$ together with a small contribution from
$K^0_L\rightarrow\pi^+\pi^-$, where the latter is known precisely from
$K^0_L$ semileptonic decays,
$A^{\bar{K}^0}_{CP}=(-0.332\pm0.006)$\%~\cite{PDG2012}. As shown in
eq.~(\ref{EQ:ACP}), the product of the two small asymmetries is
neglected. The $D^+$ decaying to the final state $K^0_S K^+$ proceeds
from $D^+\rightarrow\bar{K}^0K^+$ decay, which is SCS. In the SM,
direct $CP$ violation in SCS charmed meson decays is predicted to
occur with a non-vanishing phase that enters the diagram shown in
figure~\ref{FIG:DIAGRAMS}(b) in the Kobayashi-Maskawa
ansatz~\cite{KM}. The current average of $\Delta A_{CP}$ favors a
negative value of direct $CP$ violation in $D^0\rightarrow K^+K^-$
decay. Correspondingly, the $CP$ asymmetry in $D^+\rightarrow K^0_SK^+$
decays is more likely to have a negative value since the two $CP$
asymmetry terms shown in eq.~(\ref{EQ:ACP}) are negative. 
\section{Methodology}
We determine $A^{D^+\rightarrow K^0_S K^+}_{CP}$ by measuring the
asymmetry in the signal yield
\begin{equation}
  A^{D^+\rightarrow K^0_S K^+}_{\rm rec}=\frac
  {N_{\rm rec}^{D^+\rightarrow K^0_S K^+}-N_{\rm rec}^{D^-\rightarrow K^0_S K^-}}
  {N_{\rm rec}^{D^+\rightarrow K^0_S K^+}+N_{\rm rec}^{D^-\rightarrow K^0_S K^-}},    
  \label{EQ:ARECONI}
\end{equation}
where $N_{\rm rec}$ is the number of reconstructed decays. The
asymmetry in eq.~(\ref{EQ:ARECONI}) includes the forward-backward
asymmetry ($A_{FB}$) due to $\gamma^{*}$-$Z^0$ interference and higher
order QED effects in $e^+e^-\rightarrow
c\bar{c}$~\cite{HIGHQED1,HIGHQED2,HIGHQED3}, and the detection
efficiency asymmetry between $K^+$ and $K^-$ ($A^{K^+}_{\epsilon}$) as
well as $A_{CP}$. In addition, ref.~\cite{K0MAT} calculates another
asymmetry source, denoted $A_{\mathcal{D}}$, due to the differences in
interactions of $\bar{K}^0$ and $K^0$ mesons with the material of the
detector. Since we reconstruct the $K^0_S$ with $\pi^+\pi^-$
combinations, the $\pi^+\pi^-$ detection asymmetry cancels out for
$K^0_S$. The asymmetry of eq.~(\ref{EQ:ARECONI}) can be written as
\begin{eqnarray}
  \nonumber
  A^{D^+\rightarrow K^0_S K^+}_{\rm rec}(\cos\theta^{\rm c.m.s.}_{D^+},p^{\rm lab}_{TK^+},\cos\theta^{\rm lab}_{K^+},p^{\rm lab}_{K^0_S})
  &=&A^{D^+\rightarrow K^0_S K^+}_{CP}+A^{D^+}_{FB}(\cos\theta^{\rm c.m.s.}_{D^+})\\
  &+&A^{K^+}_{\epsilon}(p^{\rm lab}_{TK^+},\cos\theta^{\rm lab}_{K^+})+A_{\mathcal{D}}(p^{\rm lab}_{K^0_S})
  \label{EQ:ARECONII}
\end{eqnarray}
by neglecting the terms involving the product of asymmetries. In
eq.~(\ref{EQ:ARECONII}), $A^{D^+\rightarrow K^0_S K^+}_{CP}$ is the
sum of $A^{D^+\rightarrow\bar{K}^0 K^+}_{CP}$ and $A^{\bar{K}^0}_{CP}$
as stated in eq.~(\ref{EQ:ACP}), where the former is independent of
all kinematic variables while the latter is known to depend on the
$K^0_S$ decay time according to ref.~\cite{GROSSMAN_NIR}, and
$A^{D^+}_{FB}$ is an odd function of the cosine of the polar angle
$\theta^{\rm c.m.s.}_{D^+}$ of the $D^+$ momentum in the
center-of-mass system (c.m.s.). $A^{K^+}_{\epsilon}$ depends on the
transverse momentum $p^{\rm lab}_{TK^+}$ and the polar angle
$\theta^{\rm lab}_{K^+}$ of the $K^+$ in the laboratory frame
(lab). Here, $A_{\mathcal{D}}$ is a function of the lab momentum
$p^{\rm lab}_{K^0_S}$ of the $K^0_S$. To correct for
$A^{K^+}_{\epsilon}$ in eq.~(\ref{EQ:ARECONII}), we use the technique
developed in our previous publication~\cite{OLDKSPI}. We use
$D^0\rightarrow K^-\pi^+$ and $D^+_s\rightarrow\phi\pi^+$ decays where
the $\phi$ is reconstructed with $K^+K^-$ combinations and hence the
$K^+K^-$ detection asymmetry nearly cancels out~\cite{MARKO_AKK} (the
residual small effect is included in the systematic error). Since these
are Cabibbo-favored decays for which the direct $CP$ asymmetry is
expected to be negligible, in analogy to eq.~(\ref{EQ:ARECONII}),
$A^{D^0\rightarrow K^-\pi^+}_{\rm rec}$ and
$A^{D^+_s\rightarrow\phi\pi^+}_{\rm rec}$ can be written as
\begin{eqnarray}
\label{EQ:ARECD0}
  \nonumber
  A^{D^0\rightarrow K^-\pi^+}_{\rm rec}(\cos\theta^{\rm c.m.s.}_{D^0},p^{\rm lab}_{TK^-},\cos\theta^{\rm lab}_{K^-},p^{\rm lab}_{T\pi^+},\cos\theta^{\rm lab}_{\pi^+})
  &=&A^{D^0}_{FB}(\cos\theta^{\rm c.m.s.}_{D^0})+A^{K^-}_{\epsilon}(p^{\rm lab}_{TK^-},\cos\theta^{\rm lab}_{K^-})\\
  &+&A^{\pi^+}_{\epsilon}(p^{\rm lab}_{T\pi^+},\cos\theta^{\rm lab}_{\pi^+}),\\
\label{EQ:ARECDS}
  \nonumber
  A^{D^+_s\rightarrow\phi\pi^+}_{\rm rec}(\cos\theta^{\rm c.m.s.}_{D^+_s},p^{\rm lab}_{T\pi^+},\cos\theta^{\rm lab}_{\pi^+})
  &=&A^{D^+_s}_{FB}(\cos\theta^{\rm c.m.s.}_{D^+_s})\\
  &+&A^{\pi^+}_{\epsilon}(p^{\rm lab}_{T\pi^+},\cos\theta^{\rm lab}_{\pi^+}).
\end{eqnarray}
Thus, with the additional $A^{K^-}_{\epsilon}$ term in
$A^{D^0\rightarrow K^-\pi^+}_{\rm rec}$, one can measure
$A^{K^-}_{\epsilon}$ by subtracting
$A^{D^+_s\rightarrow\phi\pi^+}_{\rm rec}$ from $A^{D^0\rightarrow
  K^-\pi^+}_{\rm rec}$, assuming the same $A_{FB}$ for $D^0$ and
$D^+_s$ mesons. We also obtain $A_{\mathcal{D}}$ according to
ref.~\cite{K0MAT}. After these $A^{K^+}_{\epsilon}$ and
$A_{\mathcal{D}}$ corrections~\footnote{We define
  $A^{h^+}\equiv[N^{h^+}-N^{h^-}]/[N^{h^+}+N^{h^-}]$. Hence
  $A^{h^-}=-A^{h^+}$.}, we obtain
\begin{equation} 
  A^{D^+\rightarrow K^0_S K^+_{\rm corr}}_{\rm rec}(\cos\theta^{\rm c.m.s.}_{D^+})=A^{D^+\rightarrow K^0_S K^+}_{CP}+A^{D^+}_{FB}(\cos\theta^{\rm c.m.s.}_{D^+}).
  \label{EQ:ARECCORR}
\end{equation}
We subsequently extract $A_{CP}$ and $A_{FB}$ as a function of
$\cos\theta^{\rm c.m.s.}_{D^+}$ by taking sums and differences:
\begin{subequations} 
\begin{equation}
  \label{EQ:ACPCOS}
  A^{D^+\rightarrow K^0_S K^+}_{CP}(|\cos\theta^{\rm c.m.s.}_{D^+}|)
  =\frac{A^{D^+\rightarrow K^0_S K^+_{\rm corr}}_{\rm rec}(+\cos\theta^{\rm c.m.s.}_{D^+})    
    +A^{D^+\rightarrow K^0_S K^+_{\rm corr}}_{\rm rec}(-\cos\theta^{\rm c.m.s.}_{D^+})}{2},
\end{equation}
\begin{equation}
  \label{EQ:AFBCOS}
  A^{D^+}_{FB}(|\cos\theta^{\rm c.m.s.}_{D^+}|)
  =\frac{A^{D^+\rightarrow K^0_S K^+_{\rm corr}}_{\rm rec}(+\cos\theta^{\rm c.m.s.}_{D^+})
    -A^{D^+\rightarrow K^0_S K^+_{\rm corr}}_{\rm rec}(-\cos\theta^{\rm c.m.s.}_{D^+})}{2}.
\end{equation}
\end{subequations} 
Note that extracting $A_{CP}$ in eq.~(\ref{EQ:ARECCORR}) using
eq.~(\ref{EQ:ACPCOS}) is crucial here to cancel out the Belle
detector's asymmetric acceptance in $\cos\theta^{\rm c.m.s.}_{D^+}$.
\section{Data and event selections}
The data used in this analysis were recorded at the $\Upsilon(nS)$
resonances $(n=1,2,3,4,5)$ or near the $\Upsilon(4S)$ resonance with
the Belle detector at the $e^+e^-$ asymmetric-energy collider
KEKB~\cite{KEKB}. The data sample corresponds to an integrated
luminosity of 977 fb$^{-1}$. The Belle detector is a large solid angle
magnetic spectrometer that consists of a silicon vertex detector
(SVD), a 50-layer central drift chamber (CDC), an array of aerogel
threshold Cherenkov counters (ACC), a barrel-like arrangement of
time-of-flight scintillation counters (TOF), and an electromagnetic
calorimeter comprising CsI(Tl) crystals located inside a
superconducting solenoid coil that provides a 1.5 T magnetic field. An
iron flux return located outside the coil is instrumented to detect
$K^0_L$ mesons and to identify muons. A detailed description of the
Belle detector can be found in ref.~\cite{BELLE}.

Except for the tracks from $K^0_S$ decays we require charged tracks to
originate from the vicinity of the interaction point (IP) by limiting
the impact parameters along the beam direction ($z$-axis) and
perpendicular to it to less than 4 cm and 2 cm, respectively. All
charged tracks other than those from $K^0_S$ decays are identified as
pions or kaons by requiring the ratio of particle identification
likelihoods, $\mathcal{L}_K/(\mathcal{L}_K+\mathcal{L}_{\pi})$,
constructed using information from the CDC, TOF, and ACC, to be larger
or smaller than 0.6, respectively~\cite{BELLEPID}. For both kaons and
pions, the efficiencies and misidentification probabilities are about
90\% and 5\%, respectively.

We form $K^0_S$ candidates adopting the standard Belle $K^0_S$
criteria~\cite{BRKS}, requiring the invariant mass of the charged track
pair to be within $[0.4826, 0.5126]$ GeV/$c^2$. The ``loose'' $K^0_S$
candidates not satisfying these standard selections are also used in
this analysis with additional requirements described later.

\begin{figure}[htbp]
\centering
  \includegraphics[height=0.4\textwidth,width=0.9\textwidth]{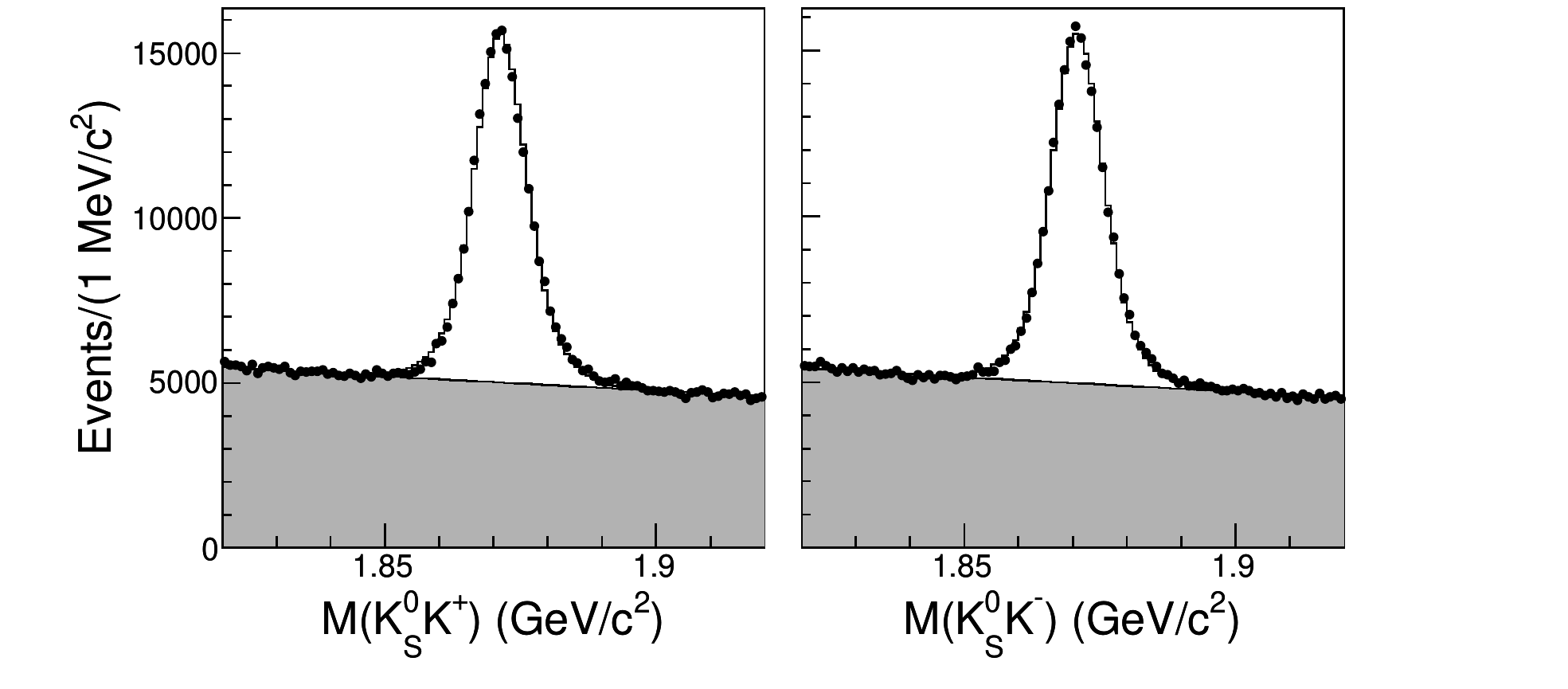}
  \caption{Distributions of $M(K^0_S K^+)$ (left) and $M(K^0_S K^-)$
    (right). Dots are the data while the histograms show the results
    of the parameterizations of the data. Open histograms represent
    the $D^{\pm}\rightarrow K^0_S K^{\pm}$ signal and shaded regions
    are combinatorial background.}
  \label{FIG:MKSK}
\end{figure}

The $K^0_S$ and $K^+$ candidates are combined to form a $D^+$
candidate by fitting their tracks to a common vertex; the $D^+$ candidate
is fitted to the independently measured IP profile to give the
production vertex. To remove combinatorial background as well as $D^+$
mesons that are produced in possibly $CP$-violating $B$ meson decays,
we require the $D^+$ meson momentum calculated in the
c.m.s. ($p^*_{D^+}$) to be greater than 2.5 and 3.0 GeV/$c$ for the
data taken at the $\Upsilon(4S)$ and $\Upsilon(5S)$ resonances,
respectively. For the data taken below $\Upsilon(4S)$, where no $B$
mesons are produced, we apply the requirement $p^*_{D^+}$$>$2.0
GeV/$c$. In addition to the selections described above, we further
optimize the signal sensitivity with four variables: the
goodness-of-fit values of the $D^+$ decay- and production-vertex fits
$\chi^2_D$ and $\chi^2_P$, the transverse momentum of the $K^+$ in the
lab $p^{\rm lab}_{TK^+}$, and the angle $\xi$ between the $D^+$
momentum vector (as reconstructed from its daughters) and the vector
joining the $D^+$ production and decay vertices. We optimize the
requirement on these four variables with the standard and loose
$K^0_S$ selections by maximizing
$\mathcal{N}_S/\sqrt{\mathcal{N}_S+\mathcal{N}_B}$, where
$\mathcal{N}_S+\mathcal{N}_B$ and $\mathcal{N}_B$ are the yields in
the $K^0_S K^+$ invariant mass signal ($[1.860, 1.884]$ GeV/$c^2$) and
sideband ($[1.843, 1.855]$ and $[1.889, 1.901]$ GeV/$c^2$) regions,
respectively. The optimal set of ($\chi^2_D$, $\chi^2_P$, $p^{\rm
  lab}_{TK^+}$, $\xi$) requirements are found to be ($<$100, $<$10,
$>$0.30 GeV/$c$, $<$40$^{\circ}$), ($<$100, $<$10, $>$0.25 GeV/$c$,
$<$115$^{\circ}$), and ($<$100, $<$10, $>$0.20 GeV/$c$,
$<$125$^{\circ}$) for the data taken below the $\Upsilon(4S)$, at the
$\Upsilon(4S)$, and at the $\Upsilon(5S)$, respectively. Note that
$p^{*}_{D^+}$ is highly correlated with $p^{\rm lab}_{TK^+}$ and
$\xi$; hence, a tighter $p^{*}_{D^+}$ requirement on the
$\Upsilon(5S)$ sample results in looser $p^{\rm lab}_{TK^+}$ and $\xi$
requirements and vice versa for the data taken below the
$\Upsilon(4S)$. The $D^+$ candidates with the loose $K^0_S$
requirement are further optimized with two additional variables: the
$\chi^2$ of the fit of tracks from the $K^0_S$ decay and the kaon from
the $D^+$ meson decay to a single vertex ($\chi^2_{Khh}$) and the
angle $\zeta$ between the $K^0_S$ momentum vector (as reconstructed
from its daughters) and the vector joining the $D^+$ and $K^0_S$ decay
vertices. The two variables are again varied simultaneously and the
optimum is found to be $\chi^2_{Khh}$$>$6 and $\zeta$$<$3$^{\circ}$
for all data. The inclusion of $D^+$ candidates with the loose $K^0_S$
requirement improves the statistical sensitivity by approximately
5\%. After the final selections described above, we find no
significant peaking backgrounds---for example,
$D^+\rightarrow\pi^+\pi^-K^+$ decays---in the Monte Carlo (MC)
simulated events~\cite{MC}. Figure~\ref{FIG:MKSK} shows the
distributions of $M(K^0_S K^+)$ and $M(K^0_S K^-)$ together with the
results of the fits described below.

Each $D^{\pm}\rightarrow K^0_S K^{\pm}$ signal is parameterized as
two Gaussian distributions with a common mean. The combinatorial
background is parameterized with the unnormalized form
$e^{\alpha+\beta M(K^0_S K^{\pm})}$, where $\alpha$ and $\beta$ are
fit parameters. The asymmetry and the sum of the $D^+$ and $D^-$
yields are directly obtained from a simultaneous fit to the $D^+$ and
$D^-$ candidate distributions. Besides the asymmetry and the sum of
the $D^+$ and $D^-$ yields, the common parameters in the simultaneous
fit are the widths of the two Gaussians and the ratio of their
amplitudes. The asymmetry and the sum of the $D^+\rightarrow K^0_SK^+$
and $D^-\rightarrow K^0_SK^-$ yields from the fit are
$(+0.048\pm0.275)\%$ and $276812\pm1156$, respectively, where the
errors are statistical.

In order to measure the $CP$ asymmetry in $D^+\rightarrow K^0_SK^+$
decays, we must also reconstruct $D^0\rightarrow K^-\pi^+$ and
$D^+_s\rightarrow\phi\pi^+$ decays: see eqs.~(\ref{EQ:ARECONII}),
(\ref{EQ:ARECD0}), and (\ref{EQ:ARECDS}). For the reconstruction of
the $D^0\rightarrow K^-\pi^+$ and $D^+_s\rightarrow\phi\pi^+$ decays,
we require the same track quality, particle identification, vertex fit
quality, and $p^{*}_{D}$ requirements as used for the reconstruction of
the $D^+\rightarrow K^0_SK^+$ decays, where the mass window for the
$\phi$ is $\pm$16 MeV/$c^2$~\cite{MARKO_AKK} of the nominal $\phi$
mass~\cite{PDG2012}.
\section{Extraction of $A_{CP}$ in the decay $D^+\rightarrow K^0_SK^+$}
\begin{figure}[htbp]
\centering
  \includegraphics[height=0.80\textwidth,width=0.75\textwidth]{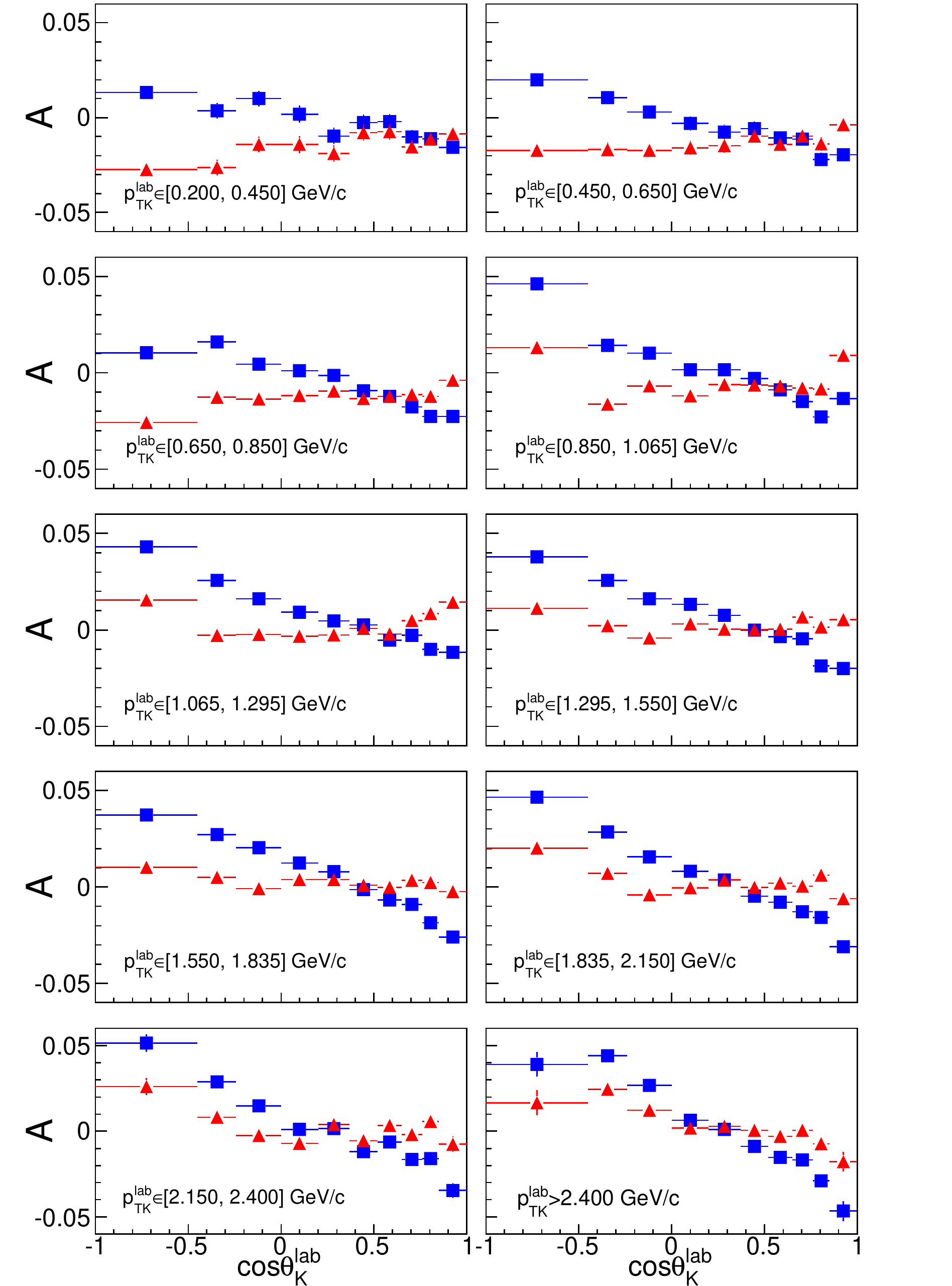}
  \caption{The $A^{K^-}_{\epsilon}$ map in bins of $p^{\rm lab}_T$ and
  $\cos\theta^{\rm lab}$ of the $K^-$ obtained with the
  $D^0\rightarrow K^-\pi^+$ and $D^+_s\rightarrow\phi\pi^+$ samples
  (triangles). The $A^{D^0\rightarrow K^-\pi^+}_{\rm rec}$ map is also
  shown (rectangles).}
\label{FIG:APIMAP}
\end{figure}

To obtain $A^{K^+}_{\epsilon}$, we first extract
$A^{D^+_s\rightarrow\phi\pi^+}_{\rm rec}$ from a simultaneous fit to
the mass distributions of $D^+_s$ and $D^-_s$ candidates with similar
parameterizations as for $D^{\pm}\rightarrow K^0_S K^{\pm}$ decays
except that, for the $D^{\pm}_s\rightarrow\phi\pi^{\pm}$ signal
description, a single Gaussian is used. The values of
$A^{D^+_s\rightarrow\phi\pi^+}_{\rm rec}$ are evaluated in
10$\times$10$\times$10 bins of the three-dimensional (3D) phase space
($p^{\rm lab}_{T\pi^+}$, $\cos\theta^{\rm lab}_{\pi^+}$,
$\cos\theta^{\rm c.m.s.}_{D^+_s}$). Each $D^0\rightarrow K^-\pi^+$ and
$\bar{D}^0\rightarrow K^+\pi^-$ candidate is then weighted with a
factor of $1-A^{D^+_s\rightarrow\phi\pi^+}_{\rm rec}$ and
$1+A^{D^+_s\rightarrow\phi\pi^+}_{\rm rec}$, respectively, in the
corresponding bin of this space. After this weighting, the asymmetry
in the $D^0\rightarrow K^-\pi^+$ decay sample becomes
$A^{K^-}_{\epsilon}$. The detector asymmetry, $A^{K^-}_{\epsilon}$, is
measured from simultaneous fits to the weighted $M(K^{\mp}\pi^{\pm})$
distributions in 10$\times$10 bins of the 2D phase space ($p^{\rm
  lab}_{TK^-}$, $\cos\theta^{\rm lab}_{K^-}$) with similar
parameterizations as used for $D^+\rightarrow K^0_S K^+$ decays except
that, for the $D^0\rightarrow K^-\pi^+$ signal description, a sum of a
Gaussian and bifurcated Gaussian is used. Figure~\ref{FIG:APIMAP}
shows the measured $A^{K^-}_{\epsilon}$ in bins of $p^{\rm
  lab}_{TK^-}$ and $\cos\theta^{\rm lab}_{K^-}$ together with
$A^{D^0\rightarrow K^-\pi^+}_{\rm rec}$ for comparison; we observe
that $A^{D^0\rightarrow K^-\pi^+}_{\rm rec}$ shows a $\cos\theta^{\rm
  lab}_{K^-}$ dependency that is inherited from $A^{D^{0}}_{FB}$ while
$A^{K^-}_{\epsilon}$ does not. The average of $A^{K^-}_{\epsilon}$
over the phase space is $(-0.150\pm0.029)\%$, where the error is due
to the limited statistics of the $D^0\rightarrow K^-\pi^+$ sample.

Based on a recent study of $A_{\mathcal{D}}$~\cite{K0MAT}, we obtain
the dilution asymmetry in bins of $K^0_S$ lab momentum. For the
present analysis, $A_{\mathcal{D}}$ is approximately 0.1\% after
integrating over the phase space of the two-body decay.

\begin{figure}[htbp]
\centering
  \includegraphics[height=0.60\textwidth,width=0.7\textwidth]{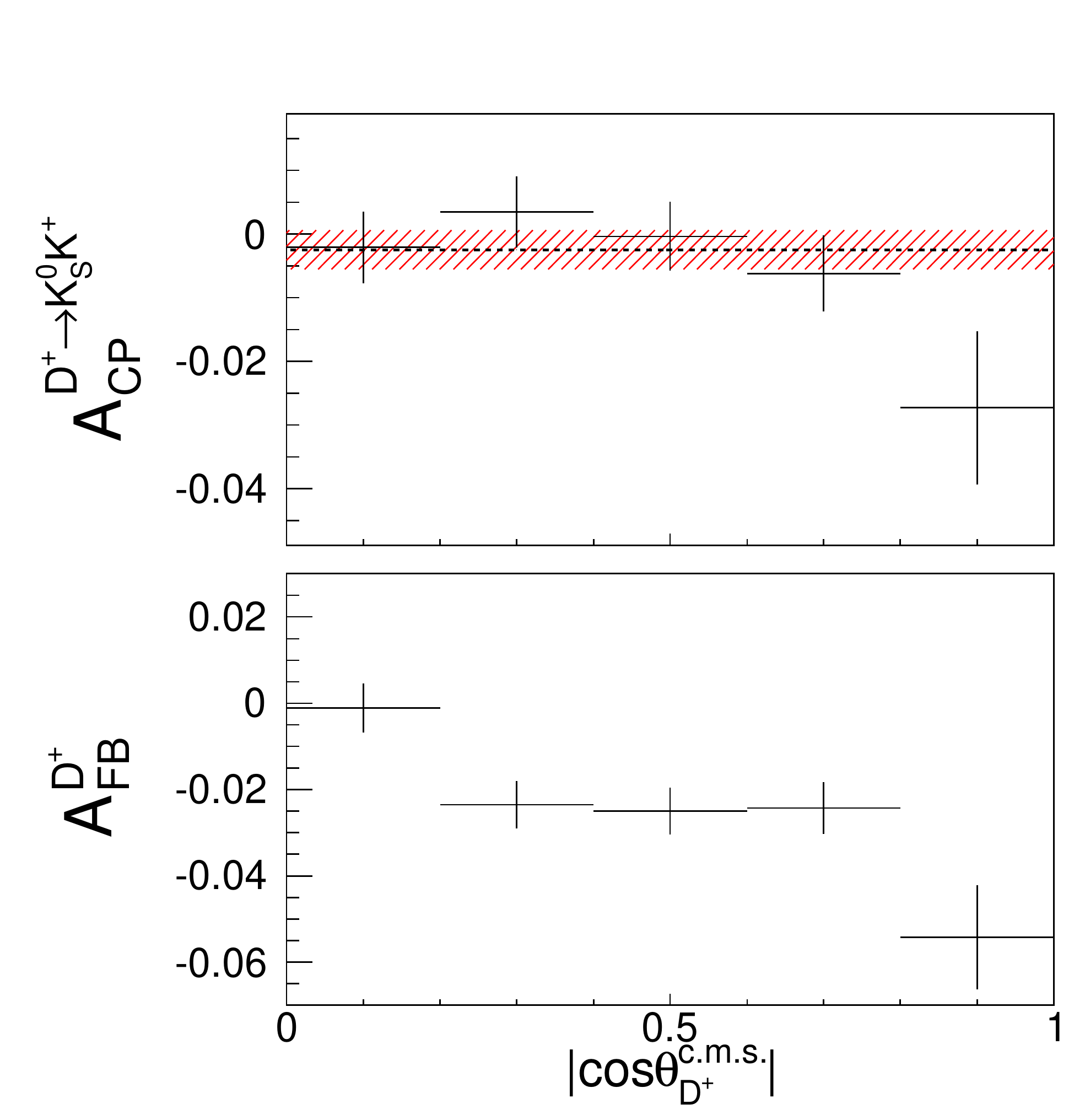}
  \caption{Measured $A_{CP}$ (top) and $A_{FB}$ (bottom) values as a function of
  $|\cos\theta^{\rm c.m.s.}_{D^{+}}|$. In the top plot, the dashed line is the
  mean value of $A_{CP}$ while the hatched band is the $\pm1\sigma_{\rm total}$
  interval, where $\sigma_{\rm total}$ is the total uncertainty.}
\label{FIG:ACP}
\end{figure}

The data samples shown in figure~\ref{FIG:MKSK} are divided into
10$\times$10$\times$16 bins of the 3D phase space ($p^{\rm lab}_{TK^+}$,
$\cos\theta^{\rm lab}_{K^+}$, $p^{\rm lab}_{K^0_S}$). Each
$D^{\pm}\rightarrow K^0_S K^{\pm}$ candidate is then weighted with a factor of
$(1\mp A^{K^+}_{\epsilon})(1\mp A_{\mathcal{D}})$ in this space. The
weighted $M(K^0_S K^{\pm})$ distributions in bins of $\cos\theta^{\rm
  c.m.s.}_{D^+}$ are fitted simultaneously to obtain the corrected
asymmetry. We fit the linear component in $\cos\theta^{\rm
  c.m.s.}_{D^{+}}$ to determine $A_{FB}$; the $A_{CP}$ component
is uniform in $\cos\theta^{\rm c.m.s.}_{D^{+}}$. Figure~\ref{FIG:ACP}
shows $A^{D^+\rightarrow K^0_S K^+}_{CP}$ and $A^{D^{+}}_{FB}$ as a
function of $|\cos\theta^{\rm c.m.s.}_{D^{+}}|$.
From a weighted average over the $|\cos\theta^{\rm c.m.s.}_{D^{+}}|$
bins, we obtain $A^{D^+\rightarrow K^0_S
  K^+}_{CP}=(-0.246\pm0.275)\%$, where the error is statistical. 
\section{Systematic uncertainty}
The entire analysis procedure is validated with fully simulated MC
events~\cite{MC} and the result is consistent with null input
asymmetry. We also consider other sources of systematic
uncertainty. The dominant one in the $A_{CP}$ measurement is the
$A^{K^+}_{\epsilon}$ determination, the uncertainty of which is mainly
due to the statistical uncertainties in the $D^0\rightarrow K^-\pi^+$
and $D^+_s\rightarrow\phi\pi^+$ samples. These are found to be 0.029\%
and 0.119\%, respectively, from a simplified simulation study. A
possible $A_{CP}$ in the $D^0\rightarrow K^-\pi^+$ final state is
estimated using
$A_{CP}=-y\sin\delta\sin\phi\sqrt{R}$~\cite{PETROV}. A calculation with
95\% upper and lower limits on $D^0-\bar{D}^0$ mixing and $CP$
violation parameters $y$, $\phi$, and strong phase difference $\delta$
and Cabibbo suppression factor $R$ from ref.~\cite{HFAG}, $A_{CP}$ in
the $D^0\rightarrow K^-\pi^+$ final state is estimated to be less than
0.005\% and this is included as one of systematic uncertainties in the
$A^{K^+}_{\epsilon}$ determination. As reported in our previous
publication~\cite{MARKO_AKK}, the magnitude of $A^{KK}_{\rm rec}$ for
the $\phi$ reconstruction in $D^+_s\rightarrow\phi\pi^+$ decays is
0.051\%, which is also added to the systematic uncertainty in the
$A^{K^+}_{\epsilon}$ measurement. By adding the contributions in
quadrature, the systematic uncertainty in the $A^{K^+}_{\epsilon}$
determination is estimated to be 0.133\%. We estimate 0.008\% and
0.021\% systematic uncertainties due to the choice of the fitting
method and that of the $\cos\theta^{\rm c.m.s.}_{D^+}$ binning,
respectively. Finally, we add the systematic uncertainty in the
$A_{\mathcal{D}}$ correction, which is 0.010\% based on
ref.~\cite{K0MAT}. The quadratic sum of the above uncertainties,
0.135\%, is taken as the total systematic uncertainty.
\section{Results}
We find $A^{D^+\rightarrow
  K^0_SK^+}_{CP}=(-0.246\pm0.275\pm0.135)\%$. This measurement
supersedes our previous determination of $A^{D^+\rightarrow
  K^0_SK^+}_{CP}$~\cite{OLDKSPI}. In Table~\ref{TABLE:SUMMARY}, we
compare all the available measurements and give their weighted
average. 

According to Grossman and Nir~\cite{GROSSMAN_NIR}, we can estimate the
experimentally measured $CP$ asymmetry induced by SM $K^0-\bar{K}^0$
mixing, $A^{\bar{K}^0}_{CP}$. The efficiency as a function of $K^0_S$
decay time in our detector is obtained from MC simulated events. The
efficiency is then used in eq.~(2.10) of ref.~\cite{GROSSMAN_NIR} to
obtain the correction factor that takes into account, for
$A^{\bar{K}^0}_{CP}$, the dependence on the kaon decay time. The
result is $0.987\pm0.007$. By multiplying the correction factor
$0.987\pm0.007$ and the asymmetry due to the neutral
kaons~\cite{PDG2012}, we find the experimentally measured
$A^{\bar{K}^0}_{CP}$ to be $(-0.328\pm0.006)\%$.
\begin{table}[htbp]
\centering
\begin{tabular}{|l|c|} \hline
Experiment                                & $A^{D^+\rightarrow K^0_S K^+}_{CP}$ (\%) \\ \hline
FOCUS~\cite{FOCUS_KSPI}                   & $+7.1\pm6.1\pm1.2$ \\ 
CLEO~\cite{CLEO_KSPI}                     & $-0.2\pm1.5\pm0.9$  \\
Belle (this measurement)                  & $-0.246\pm0.275\pm0.135$ \\ \hline
New world average                         & $-0.23\pm0.30$  \\ \hline
\end{tabular}     
\caption{Summary of $A_{CP}^{D^+\rightarrow K^0_S K^+}$ measurements (where the
  first uncertainties are statistical and the second systematic), together with
  their average (assuming the uncertainties to be uncorrelated, the
  error on the average represents the total uncertainty).}
\label{TABLE:SUMMARY}
\end{table}
\section{Conclusion}\label{SEC:CONCLUSION}
We report the most sensitive $CP$ asymmetry measurement to date for
the decay $D^+\rightarrow K^0_S K^+$ using a data sample corresponding
to an integrated luminosity of 977 fb$^{-1}$ collected with the Belle
detector. The $CP$ asymmetry in the decay is measured to be
$(-0.25\pm0.28\pm0.14)\%$. After subtracting the contribution due to
$K^0-\bar{K}^0$ mixing ($A^{\bar{K}^0}_{CP}$), the $CP$ asymmetry in
the charm decay ($A^{D^+\rightarrow\bar{K}^0 K^+}_{CP}$) is measured
to be $(+0.08\pm0.28\pm0.14)\%$, which can be compared with direct
$CP$ violation in $D^0\rightarrow K^+K^-$. For the latter the current
averages of $\Delta A_{CP}$ and $CP$ asymmetry in $D^0\rightarrow
K^+K^-$ favor a negative value~\cite{HFAG}. Our result, on the other
hand, does not show this tendency for $D^+\rightarrow\bar{K}^0K^+$
decays, albeit with a significant statistical uncertainty.
\acknowledgments
We thank the KEKB group for the excellent operation of the
accelerator; the KEK cryogenics group for the efficient operation of
the solenoid; and the KEK computer group, the National Institute of
Informatics, and the PNNL/EMSL computing group for valuable computing
and SINET4 network support.  We acknowledge support from the Ministry
of Education, Culture, Sports, Science, and Technology (MEXT) of
Japan, the Japan Society for the Promotion of Science (JSPS), and the
Tau-Lepton Physics Research Center of Nagoya University; the
Australian Research Council and the Australian Department of Industry,
Innovation, Science and Research; the National Natural Science
Foundation of China under contract No.~10575109, 10775142, 10875115
and 10825524; the Ministry of Education, Youth and Sports of the Czech
Republic under contract No.~LA10033 and MSM0021620859; the Department
of Science and Technology of India; the Istituto Nazionale di Fisica
Nucleare of Italy; the BK21 and WCU program of the Ministry Education
Science and Technology, National Research Foundation of Korea Grant
No.~2011-0029457, 2012-0008143, 2012R1A1A2008330, BRL program under
NRF Grant No. KRF-2011-0020333, and GSDC of the Korea Institute of
Science and Technology Information; the Polish Ministry of Science and
Higher Education and the National Science Center; the Ministry of
Education and Science of the Russian Federation and the Russian
Federal Agency for Atomic Energy; the Slovenian Research Agency;  the
Swiss National Science Foundation; the National Science Council and
the Ministry of Education of Taiwan; and the U.S. Department of Energy
and the National Science Foundation. This work is supported by a
Grant-in-Aid from MEXT for  Science Research in a Priority Area (``New
Development of Flavor Physics''), and from JSPS for Creative
Scientific Research (``Evolution of Tau-lepton Physics''). B.~R.~Ko
acknowledges support by NRF Grant No. 2012-0007319, and E.~Won by NRF
Grant No. 2010-0021174.

\end{document}